\documentclass[aps,preprint,preprintnumbers,nofootinbib,showpacs]{revtex4}

\usepackage{amsmath}
\usepackage{amssymb}
\usepackage{mathtools}
\usepackage{dsfont}
\usepackage{color}
\usepackage{graphicx,accents}
\usepackage[normalem]{ulem}

\definecolor{indred}{rgb}{0.8, 0.36, 0.36}
\def\bea{\begin{eqnarray}}
\def\eea{\end{eqnarray}}
\def\sea{\nonumber \\&&}
\def\lla{\left\langle}
\def\rra{\right\rangle}

\def\zb{\beta}

\newcommand{\bra}[1]{\lla#1\right|}
\newcommand{\ket}[1]{\left|#1\rra}

\def\ssc{\scriptscriptstyle}
\def\lsim{\mathrel{\raise.3ex\hbox{$<$\kern-.75em\lower1ex\hbox{$\sim$}}} }
\def\gsim{\mathrel{\raise.3ex\hbox{$>$\kern-.75em\lower1ex\hbox{$\sim$}}} }
\makeatletter
\DeclareRobustCommand{\cev}[1]{%
  \mathpalette\do@cev{#1}%
}
\newcommand{\do@cev}[2]{%
  \fix@cev{#1}{+}%
  \reflectbox{$\m@th#1\vec{\reflectbox{$\fix@cev{#1}{-}\m@th#1#2\fix@cev{#1}{+}$}}$}%
  \fix@cev{#1}{-}%
}
\newcommand{\fix@cev}[2]{%
  \ifx#1\displaystyle
    \mkern#23mu
  \else
    \ifx#1\textstyle
      \mkern#23mu
    \else
      \ifx#1\scriptstyle
        \mkern#22mu
      \else
        \mkern#22mu
      \fi
    \fi
  \fi
}

\makeatother
\begin{document}
\preprint{\vbox{\hbox{NCU-HEP-k103}
\hbox{May 2024}
}}


\title{\boldmath  Noncommutative Number Systems for Quantum Information 
}

\author{Otto C. W. Kong}  

\affiliation{ Department of Physics and Center for High Energy and High Field Physics,
National Central University, Chung-li, Taiwan 32054  \\
}


\begin{abstract}
\vspace*{.1in}
Dirac talked about q-numbers versus c-numbers. Quantum observables are q-number variables
that generally do not commute among themselves. He was proposing to have a generalized
form of numbers as elements of a noncommutative algebra. That was Dirac's appreciation of the 
mathematical properties of the physical quantities as presented in Heisenberg's new quantum
theory. After all, the familiar real, or complex, number system only came into existence through 
the history of mathematics. Values of physical quantities having a commutative product is an 
assumption that is not compatible with quantum physics. The revolutionary idea of Heisenberg 
and Dirac was pulled back to a much more conservative setting by the work of Schr\"odinger, 
followed by Born and Bohr. What Bohr missed is that the real number values we obtained from 
our measurements are only a consequence of the design of the kind of experiments and our 
using real numbers to calibrate the output scales of our apparatus. It is only our modeling of the 
information obtained about the physical quantities rather than what Nature dictates. We have 
proposed an explicit notion of definite noncommutative values of observables that gives a picture 
of quantum mechanics as realistic as the classical theory. In this article, we illustrate how matrices
can be taken as noncommutative (q-)numbers serving as the values of physical quantities, each 
to be seen as a piece of quantum information. Our main task is to clarify the subtle issues involved 
in setting up a conventional scheme assigning matrices as values to the physical quantities.
\end{abstract}

\maketitle

\begin{center}
{\em To be Is to be the value of a variable.} \hspace*{.5in} --- {\em V. Quine}
\end{center}

\section{Introduction}
To think about information through the perspective of  Quine's dictum, we can see that a piece
of information is practically always encoded in a physical system --- yes, essentially as the value
of an observable. A piece of classical information is generally represented by a real number, 
essentially as the value of a classical observable. Now, what about a piece of quantum information? 
First of all, the nature of it as more than, or at least not, a piece of classical information says that 
it is not supposed to be possible to be represented by a single real number. Thinking about that 
as encoded in a physical quantity under the quantum theory, the quantity carries the quantum
information and there is no reason to see that as its value. A quantum observable certainly
cannot have a definite (real number) value, in general. The only exception may be the case when 
one has an eigenstate of a particular observable, for which the values of other observables not 
commuting with the latter certainly have no definite values of the kind. Even in that case, however,
it is not difficult to appreciate then that representing the information carried by the eigenstate for 
the particular observable simply as a piece of classical information may be incomplete.  For
example, we know it has zero Heisenberg uncertainty and that is some extra information. 
Heisenberg's revolutionary idea for the theory was exactly to give up the classical idea that
physical quantities have to be real-valued variables; Dirac called them q-number-valued variables. 

In the history of number theory, more and more general notions of numbers have been introduced
as values of variables, especially as solutions to algebraic equations of one or more variables.
Irrational numbers started from the introduction of the otherwise non-existing solution to the 
equation $x^2-2=0$. Complex numbers started from the introduction of the otherwise non-existing 
solution to the equation $x^2+1=0$. The equation $xy-1=0$ has no solution within the system of 
natural numbers or integers. Yet, it has an infinite family of solutions within the system of rational 
numbers, and beyond. What about an equation like $xy -yx -1=0$?  The equation as written 
indicates that  it cannot be consistent with a commutative product between the variables $x$ and 
$y$. As such, solutions certainly do not exist within any system of numbers with a commutative 
product. Following the tradition in mathematics, it is not unreasonable to introduce a more general 
notion of numbers as solutions to the equation. Now, here is the celebrated equation in 
quantum  mechanics
 \bea\label{h}
 x p - p x -i\hbar = 0 \;,
 \eea
the Heisenberg commutation relation which essentially defines the theory. The position and
momentum observables as dynamic variables can only have (definite) values, mathematically 
as solutions to the above equation, within a new number system that generalizes the complex 
numbers and have generally noncommutative products. Instead of seeing quantum observables 
as not having definite values and using notions like Born probability to create a mysterious and
controversial picture of quantum physics, introducing noncommutative numbers for the
modeling of values of physical quantities may be more reasonable. That was the wisdom of
Dirac.

As Dirac picked up on the new theory of Heisenberg in 1925, he came to clarify the mathematical
properties of Heisenberg's new way of dealing with physical quantities. In Dirac's second paper
 \cite{D2}, he stated that ``To distinguish the two kinds of numbers, we shall call the quantum
variables q-numbers and the numbers of classical mathematics which satisfy the commutative law 
c-numbers, while the word number alone will be used to denote either a q-number or a c-number."
He also said that ``At present one can form no picture of what a q-number is like." This article 
takes up the task of presenting Dirac's q-numbers. Noncommutative numbers may be a more
appropriate name.

Dirac was more explicit about the q-numbers as values of physical quantities in a later paper
 \cite{D7}:`` Owing to the fact that we count the time as a c-number,  we are 
 allowed to use the notion of the value of the dynamical variable at any instance of time. This 
 value is a q-number, capable of being represented by a generalized `matrix', \dots". 
 He pointed at `matrices' as the candidates. It is mathematically clear that matrices
 are natural candidates to be considered as some noncommutative generalization of numbers.
 Yet, the usual picture is that the quantum observables themselves, rather than their values
 for any fixed state, are matrices. Takesaki, in a preface to his three-volume book on the 
 Theory of Operator Algebra \cite{T}, suggested the theory is a kind of number theory, beyond 
 ``Integers, rational numbers, real numbers and complex numbers" that `` are all commutative."  
In resonance to Dirac, he stated that ``The advent of quantum physics at the turn of century 
forced one to consider non-commutative variables. One needed to broaden the concept of 
numbers." And that is exactly what we said above. Takesaki had not presented any 
noncommutative number system explicitly either. Nor has he addressed the relation between 
a quantum observable and its various noncommutative number values in relation. Noted that 
some other physicists, including notably t'Hooft, have been talking about `noncommutative 
numbers' somewhat. Yet, to the best of our knowledge, the general situation remains the same.
There has been no explicit presentation of any system of noncommutative numbers for modeling 
the values of physical quantities. We consider the latter the key issue distinguishing a general 
noncommutative algebra from what may be justified to be called a system of noncommutative 
numbers. That may be what has been missing in the thinking of most mathematicians against 
Takesaki's insight in using the term noncommutative number system among various 
noncommutative algebras.  
 
To give a realistic picture to quantum physics similar to that of the classical theory, 
we have introduced mathematically explicit notions of the noncommutative values of 
quantum observables in recent years \cite{088,079}. The approach was first inspired by 
the mathematical idea of noncommutative geometry \cite{C}, and the author's discomfort
about the assumption of the classical model of physical space as remaining valid for 
the quantum theory while the position coordinate observables are elements of 
a noncommutative algebra of quantum observables, at the great expense of an object 
not having a definite position. Why would we be  happy to accept such a counter-intuitive 
notion without making an effort to find a better mathematical model for the notions of 
positions and physical space that may describe the quantum positions definitely? As 
Takesaki also said : ``\dots, that in practice we never observe points directly only 
approximately by successive evaluations of coordinates." \cite{T}. From the point of view 
of theoretical physics, a theory describes relations among physical quantities as variables 
all with specific values to be determined in various specific experimental settings. What 
we are doing with our experiments is basically extracting information about a physical
system. A successful theory should describe all the available information, classical or
quantum, within its framework. The value of a particular physical quantity should then
be taken as the complete information about it as experimental available for each specific 
state. Whether one can get it out of a single simple reading of the output from the apparatus,
however, does not matter. The latter also depends very much on our technology and skills,
including the design of the apparatus and the experiment. Our experimental verification of 
a theory relies on checking the relation among those values. Yet, the exact values are only
defined in convention to the choice of the frame of reference and physical units. With the
usual projective measurement, a single eigenvalue outcome obtained hardly gives any
useful information. But by repeating the procedure enough, one can get the full statistical
distribution of that as predicted by the theory with no restriction on the precision 
achievable in principle. The amount of information involved can be seen as $N-1$ pieces 
of classical information, as the relative frequencies of the $N$ possible outcomes. 
Assuming no eigenstate degeneracy for the observable, $N$ would be the dimension of
the Hilbert space of states. A better representation of that can be given by the sequence 
of moments of the distribution. The Heisenberg uncertainty is a number among them. 
The full information the theory gives for an observable is, however, beyond that. Vogt
 \cite{V}, for example, illustrates that those distributions of position and momentum 
 observables do not uniquely determine the state, hence not carrying full information
 about it. We have recently discussed how the noncommutative values of the basic local
 observables, however, carry complete information about a quantum system independent
 of whether it is in an entangled state \cite{097}. Our first picture of the noncommutative 
 value can be represented as a sequence of complex numbers involving twice the number 
 of real numbers as the dimension of the Hilbert space. The key feature is that the 
 noncommutative algebra of such values, here the sequences, for the observables, 
 altogether forms a state-dependent homomorphic image of the observable algebra. 
 Constructing a sensible picture of such noncommutative values is about finding a way 
 to look at a state as essentially such a homomorphism. The
standard mathematical definition of a state, for an algebra, is that of a functional. The 
latter gives a complex number value for each element of the algebra. Applying to classical 
observable algebra in physics, the functional is exactly an algebraic homomorphism 
preserving the algebraic relations among the observables as dynamical variables in their 
values. Without that, one cannot check the predictions of the theory experimentally. 
For the quantum case, we need something beyond the value of that functional. Our 
first picture of the noncommutative value here discussed is essentially the sequence of 
coefficients for the Taylor series expansion of the expectation value as a function of the 
states. It naturally generalizes the real/complex number values of a state functional to
an element of a noncommutative algebra as an isomorphic images of the observable
algebra. 

The noncommutative values we talk about in the last paragraph fail to be noncommutative 
numbers as we do not have a description of the algebra of such values, particularly the 
product expression, independent of the state that defines them. The product between such 
noncommutative values of observables for different states then cannot be defined. We still
talk about them here because we consider the physics issues about the notion of 
noncommutative values of quantum observables easier to appreciate from the analyses. 
At least for the author, it served as an important background to appreciate what we are after 
here, the noncommutative numbers as an alternative description of such values. For each 
state, the former algebra of noncommutative values of observables is isomorphic to the 
algebra of their noncommutative number values. That is to say, it is an alternative description 
of what is the same algebra at the abstract level in a state-independent manner. It is 
a single mathematical system that serves as the noncommutative, q-number, values of
all observables in any state. Our candidate for the noncommutative numbers is the new 
notion of DH matrix values \cite{097} to be discussed in the next section, followed by 
a section on a detailed presentation of the system of $2\times 2$ matrices as one of the 
noncommutative number values for observables of a qubit. That aims to illustrate all the 
tricky issues lying in the way of understanding their relations to the quantum observables 
explicitly. The last section concludes the article.

 \section{DH Matrix Values for Observables as Noncommutative Number Values}
 In a study of quantum information transmission in a system of qubits in the Heisenberg 
 picture \cite{DH}, Deutsch and Hayden introduced their {\em descriptors} as local
 carriers of the complete information about the system. A sketch of the reasoning is
 as follows: The state is fixed in the Heisenberg picture and can be taken as any
 convenient reference state denoted as $\ket{0}$. All the information for the quantum 
 system is then encoded in the observables that evolve with time. Taking an 
 independent set of local basic observables, their time-evolved matrix representation,
 always contains the complete information of the system. For each qubit, the basic
 observables can be taken as (two out of) $\sigma^{\ssc 1}$,  $\sigma^{\ssc 2}$,
 and  $\sigma^{\ssc 3}$. For a particle system, they are naturally the position and
 momentum observables of the particles. The completeness of the information is 
 a consequence of the fact that all observables, including the nonlocal ones, can be 
 expressed algebraically in terms of the local basic observables. 
 
 We have pressented a careful analysis of the idea of Deutsch and Hayden and its 
 important implications on the fascinating issue of the (non)locality of quantum systems 
 in Ref.\cite{097}. In the latter paper, we adapted the key idea into a direct definition of
 a noncommutative value of a quantum observable as the DH matrix value, for the 
 analysis of the locality issue (see also Refs.\cite{HV,Th}). In this article, we focus only
 on the subject matter of seeing the DH matrix values as the noncommutative number
 values of quantum observables and do not have to bother about if the quantum 
 system is a composite one or not.  
 
 To recapitulate,  one starts with a reference state.  For any state $\ket\phi$ (with
  $\lla\phi|\phi\rra=1$), we can find a unitary transformation $U_{\!\ssc\phi}$ that takes the 
  reference state to $\ket\phi$. For the expectation value for any observable $\zb$, we have 
 \bea\label{dh}
 \bra\phi\zb\ket\phi = \bra{0} U_{\!\ssc\phi} ^\dag \zb U_{\!\ssc\phi} \ket{0}\;.
 \eea
 The idea of Deutsch and Hayden \cite{DH}, distilled out of the dynamical setting, can be
 appreciated essentially as to take the matrix of $U_{\!\ssc\phi} ^\dag \zb U_{\!\ssc\phi}$ as 
 the value of the observable $\zb$, that is our DH matrix value \cite{097}. We are most 
 thinking about a quantum system with a Hilbert space of finite dimensions here, with $\zb$ 
 and $U_{\!\ssc\phi}$ given as explicit matrices, but generalizing to the countable infinite 
 dimension case for particle systems is in principle not a problem. The notion of DH matrix 
 values clearly satisfies our criteria above for noncommutative values of observables. For 
 a fixed  $U_{\!\ssc\phi}$, we have
 \bea
 [\cdot]_{\ssc U_{\!\ssc\phi}} : \quad \zb \quad \mapsto \quad U_{\!\ssc\phi} ^\dag \zb U_{\!\ssc\phi} 
 \eea
 as a state-dependent evaluational homomorphism on the observable algebra. The map is 
 in fact one-to-one, as the zero matrix as the additive identity in the algebra cannot be 
 the value of any observables other than itself. A DH matrix value in itself is simply a (square)
 matrix. As such, the multiplication among such values is straightforward mathematically. 
 One can then take the system of $N\times N$ matrices as a system of noncommutative 
 numbers to be used to model the values of physical quantities in the quantum theory of 
 the physical system, more or less as what Dirac had kind of spelled out. That is to say, 
 a state $\ket\phi$ as a normalized vector in an $N$-dimensional Hilbert space can be seen 
 as an evaluative homomorphism  $[\cdot]_{\ssc U_{\!\ssc\phi}}$ that assigns such a 
 noncommutative number as an  $N\times N$ matrix to each observable. 
 
 A vigilant reader may raise the concern that Eq.(\ref{dh}) does not fix $U_{\!\ssc\phi}$. The 
 choice of an admissible $U_{\!\ssc\phi}$ is essentially a notion of choice of frame of reference.
  The issue is also related to the explicit matrix representation of an observable in quantum
  mechanics. The thing is we have been so accustomed to the latter that we may forget an
  observables should be rather a dynamic variable the value of which is different for different 
  states and general changes with time. Observables and their values being given through the
  same kind of mathematical quantity is certainly confusing. An observable is not a number, nor
  a value itself. But again the value of an observable is only defined based on a full system of
  reference which includes a frame of reference and at least one conventional defined reference 
  value. What is physical is the mathematical relations of the different values of the same kind, 
  {\em i.e} quantities with the same physical dimension. As the existing reference values or chosen
  standards or the fundamental units such as what is a meter of length are all classical notions,
  extending the use to the quantum noncommutative values is a tricky business. At this point,
  we choose to focus first only on the theoretical and mathematical side of the question, 
  leaving any practical issues to be addressed later. For the case of the observable algebra
  being one of  $N\times N$ matrix variables, the choice of reference frame is about fixing
  an exact basis for the Hilbert space or an explicit description of the system phase space. We
 have advocated the idea that physical frames of reference, and the related notion of relativity 
 symmetry should firstly be one for the phase space \cite{095,096}. The physical picture of 
 space(time) comes only from the configuration space of the one-particle phase space of a 
 successful dynamical theory. Moreover, from that (relativity) symmetry point of view, the 
 quantum phase space as an irreducible representation cannot be broken down into independent 
 parts of the configuration and the momentum spaces. That is a parallel to the case of the 
 (classical) relativity symmetry in the `relativistic' case cannot have space and time as 
 representations but only one spacetime. The conclusion is that the mathematical model for the 
 phase space is the model for the spacetime for the quantum case \cite{088}. Next, we need the 
 standard reference values. In the classical case, it is only about the physical definition of each
 fundamental unit, such as what is one meter and what is one gram. However, the classical
 standard is all about classical, real number values. One can think about such physical quantities 
 as noncommutative values. But they remain ones that can be well approximated by using real 
 numbers. The state of the reference system with an observable bearing the standard value is 
 taken classically, or a practical projective measurement is actually involved. For the quantum 
 case at hand, we need to give at least one standard DH matrix value from a practical physical 
 observable in a definite quantum state. That is what is given as the matrix representation of 
 the observable $\zb$. The DH matrix value is $U_{\!\ssc\phi} ^\dag \zb U_{\!\ssc\phi}$. When 
 $ U_{\!\ssc\phi}$ is the identity matrix, that is just $\zb$ itself, not as a dynamical variable, but 
 as its DH matrix value for the reference state $\ket{0}$ that is conventionally defined. The 
 latter has to be done in parallel with the physically defined basis of the Hilbert space including 
 $\ket{0}$ as one among the basis vectors. The question is beyond the classical notion of
 the different physical units, do we have to take more conventional defined DH matrix values
 to mathematically unambiguously defined all such values? 
  
  In classical physics, we have one conventionally defined value for each physical quantity 
  in a fundamental physical unit. Adopting those to the case of the quantum noncommutative
  values, it does not look like the noncommutativity itself necessarily leads to extra 
  complication. After all, fundamental physical properties such as the mass and charge of
  a particle are still taken as real number quantities. To look at the problem another way,
  however, we may need more than one conventionally defined DH matrix value even for 
  quantities having the same physical units. One can think about having a consistent matrix
  representation of the observable algebra for a system. One can assign explicit matrices
  to a set of Hermitian generators of the algebra, the rest of the elements must then have
  their matrix representations exactly as obtained from their algebraic expressions in terms
  of the generators. Such a set of generators is exactly the notion of the (local) basic 
  observables as discussed in Ref.\cite{097}. For a system of qubits, we have two for each 
  qubit. For a system of spin-zero particles, we have the position and momentum observables.
  
  We have tried to present a careful illustration of how to look at a specific $N\times N$ matrix
 as the DH matrix value of an observable and hence a noncommutative number. That requires 
 conventionally and consistently defined DH matrix values for a set of algebraically independent 
 observables, as the (local) basic observables. The latter can be given by the matrix 
 representations of those observables commonly used in the theory though without seeing them 
 explicitly as reference values for the observables rather than the observables themselves. Again 
 the latter should be dynamical, DH matrix-valued, variables. The evalutational map from the
 observable algebra to the value being one-to-one allows such a mathematical description of 
 the observables. Yet, they became, for many of us, somewhat of a source of confusion for 
 appreciating the matrices as values rather than observables. Taking a matrix representation
 of the observables is really taking that as their conventionally defined DH-matrix values for
 the reference state. Under that picture, before a generic state $\ket\phi$ is specified, it is
 $U_{\!\ssc\phi}$ as a variable that gives $U_{\!\ssc\phi} ^\dag \zb U_{\!\ssc\phi}$ as the
 dynamic variable $\hat\zb$ with the matrix $\zb$ as the reference value. Specifying the state
 specifies  $U_{\!\ssc\phi}$ and hence the value of $U_{\!\ssc\phi} ^\dag \zb U_{\!\ssc\phi}$.
 In the section below, we give an explicit illustration of all that for the simple case of a qubit, 
 $N=2$. The example would help readers to better appreciate what we discussed here in this 
 section. 
  
  Mathematically, it is easy to see that the algebra of $N\times N$ matrices can be seen as
  a system of noncommutative numbers, and such systems can be embedded into another
  of a bigger $N$. Taking $N$ to infinity is not difficult to consider either, mathematically. 
  Then, one can entertain the idea of having one single system of noncommutative numbers. 
  Note that the dimension of the Hilbert space for a particle is the countable infinity, despite
  what a naive thinking about the Schr\'odinger representation may suggest \cite{DHS}.

\section{A Simple Noncommutative Number System Qubit Observables Take Values In}
Let us use the simple example of a qubit to look at an illustrative example of a noncommutative
number system, that of $2\times 2$ matrices, and check explicitly the issues involved in having 
them as noncommutative (DH matrix) values of the observables. There are only two independent
physical observables mathematically. We use three variables $s^i$, $i=1 \to 3$ satisfying
$s^i s^j=\delta^{ij} I + i \epsilon^{ij}_{\ k} s^k$. (with the Einstein summation convention). 
A familiar picture is that of a system of spin-half with $\frac{\hbar}{2} s^i$ as the three spin 
components. Their standard matrix representation is given by the Pauli matrices $\sigma^i$. 
As two among the latter can be taken as generators for the full algebra of $2\times 2$ matrices.
Any dynamic variable $\zb$ can be expressed theoretically as a quadratic polynomial of the two  
chosen $s^i$, times an appropriate physical unit. It can also be expressed more conveniently 
as the complex linear combination $\zb= a_\mu s^\mu$, where we have $\mu= 0 \to 3$ with 
$s^{\ssc 0} \equiv I$, a constant observable. Physical, Hermitian, ones are the ones with $a^\mu$ 
being real numbers. Yet, we need the non-Hermitian ones to complete the algebra. The reference 
state $\ket{0}$ is the $s^{\ssc 3}$ eigenstate with eigenvalue $+1$. A generic qubit state can be
parametrized as 
$ \left|\phi\rra= {c}   \left|0\rra + {s}  \left|1\rra$, ${c}\equiv\cos(\frac{\theta}{2}) e^{\frac{-i\psi}{2}}$ 
and ${s}\equiv \sin(\frac{\theta}{2}) e^{\frac{i\psi}{2}}$, $0 \leq \theta \leq\pi$, $0 \leq \psi < 2\pi$. 
From Eq.(\ref{dh}) a simple choice of $U_{\!\ssc\phi}$ is given by
  \bea
  U_{\!\ssc \phi} =\left(\begin{array}{cc}
  c & - \bar{s} \\
  s & \bar{c} 
  \end{array}  \right) ,
  \eea
 with the other basis vector $\ket{1}$ as the $s^{\ssc 3}$ eigenstate with eigenvalue $-1$. 
 It is this particular orthonormal basis of the Hilbert space that gives the standard matrix 
 representation, which reads $a_\mu \sigma^\mu$ for $\zb$. The DH matrix value of an 
 observable $\zb$ for the state $\ket\phi$ is then given by
 \bea&&
 [\zb]_{\!\ssc \phi} = U^{-1}_{\!\ssc \phi}  a_\mu \sigma^\mu U _{\!\ssc \phi} 
 \sea\hspace*{.2in}
 = a_{\ssc 0}  \sigma^{\ssc 0} + ( a_{\ssc 3} + a_{\ssc 1}SC' +a_{\ssc 2}SS')  \sigma^{\ssc 3} 
  + (a_{\ssc 3}S +a_{\ssc 1}CC' +a_{\ssc 2}CS') \sigma^{\ssc 1}  - (a_{\ssc 1}S'-a_{\ssc 2}C') \sigma^{\ssc 2},
 \label{v}\eea 
 where we have introduced $C=\cos\!\theta$, $S=\sin\!\theta$, $C'= \cos\!\psi$, and $S'= \sin\!\psi$;
 $\sigma^{\ssc 0}\equiv s^{\ssc 0}$. Note that any given observable, as a dynamic variable, has 
 a value that depends only on the two real variables $\theta$ and $\psi$ which describe the state. 
 The latter assigns definite DH matrix values to all observables. Again, the DH matrix values of the 
 observables for any particular state maintain among them all the algebraic relations among the 
 observables. Having explicit knowledge about the state or not, knowing its DH matrix values 
 for the basic observables $s^{\ssc 3}$ and $s^{\ssc 1}$ allows us to determine theoretically
 such values of all observables. Of course, that also fixes the exact state. 
 
The above is all straightforward. Yet, we want to look carefully at the not-so-obvious subtleties. 
 Firstly, a generic choice of $U_{\!\ssc\phi}$ solving $\ket\phi= U_{\!\ssc\phi} \ket{0}$ is given by 
   \bea    U_{\!\ssc\phi_\chi} =
\left(\begin{array}{cc}
   c & - \bar{s} \\
   s & \bar{c} 
   \end{array}  \right)
   \left(\begin{array}{cc}
      1 & 0 \\   0 & e^{i\chi} 
      \end{array}  \right) \;.
   \eea
It takes $\ket{1}$ to $e^{i\chi}( -\bar{s}  \ket{0} +\bar{c} \ket{1})$. As the above expression
suggests, it is a product of two unitary matrices, our particularly chosen $U_{\!\ssc\phi}$ 
and that of $U_\chi = ${\tiny $\left(\begin{array}{cc}
      1 & 0 \\   0 & e^{i\chi} 
      \end{array}  \right)$}.  
The latter unitary transformation takes $\ket{1}$ to $\ket{1_\chi} \equiv e^{i\chi}\ket{1}$.  So, 
picking any nontrivial value of $\chi$ is equivalent to having the DH matrix value, hence also 
the matrix representation of the observables, taken with a somewhat different basis. $\ket{1}$ 
and $\ket{1_\chi}$ represent the same physical state. But the different phase factor relative to
the other basis vector $\ket{0}$ has implications.  $\ket{\phi_\chi}= {c} \left|0\rra + {s}  \left|1_\chi\rra$
is not the same state as  $ \left|\phi\rra= {c}  \left|0\rra + {s}  \left|1\rra$. Taking $\ket{0}$ and 
$\ket{1}$ as our basis and taking $\ket{0}$ and  $\ket{1_\chi}$ instead are two different choices 
of frames of reference for the Hilbert space as the phase space, though the reference state
$\ket{0}$ has the same coordinates as given by $\theta=0=\psi$. To look at that in another 
way, $U_{\!\ssc\phi_\chi}$ for the reference state is then given by  $U_\chi$, in conflict with
our conventional rule of having the matrix representations of the observables correspond to 
their DH matrix values for the reference state, as conventionally chosen. Alternatively, one
can interpret that as taking a different matrix representation of the observable algebra as 
$U_{\chi}^{-1} \sigma^{\ssc 1} U_{\chi}$ is 
{\tiny $\left(\begin{array}{cc}
      0 & e^{i\chi} \\  e^{-i\chi} & 0
      \end{array}  \right)$} 
instead of $\sigma^{\ssc 1}$. There is nothing wrong about taking the matrix with any
chosen nonzero value of $\chi$ to be the representation of $s^{\ssc 1}$.  A generic matrix 
representation of an abstract algebra is exactly a homomorphism, in line with our 
noncommutative value picture. When one picks a representation of a set of generators for
the algebra, that defines a representation for the whole algebra. The extra $U_\chi$ is
a unitary transformation that takes the representation of  $\zb= a_\mu \sigma^\mu$ to
the unitarily equivalent representation of $\zb= U_\chi^{-1} a_\mu \sigma^\mu U_\chi$,
though $\sigma^{\ssc 3}$ happens to be invariant under the transformation. 
Similarly, the DH matrix values of all observables for a state $\ket\phi$ can be seen as
another such representation with $\ket\phi$ as the reference state, and $-\bar{s}  \ket{0} +\bar{c} \ket{1}$
as the other basis vector to complete the reference frame. Hence, when we have a 
conventional chosen matrix representation of a maximal set of independent observables 
as generators for our observable algebra, there is one unique form of $U_{\!\ssc\phi}$ for 
a parametrized form of a generic state $\ket\phi$ that gives the DH matrix values of any 
observables for any state of the system as in Eq.(\ref{v}). 

Before a state is specified, an observable like a component of spin is a dynamic variable.
It is the physical concept of a quantity as an abstract notion. It has no `numerical' value. 
When we represent the dynamic variables in the quantum theory by a matrix, such as taking
$s^{\ssc 3}$ as the matrix $\sigma^{\ssc 3}$, the latter, of course, comes from  its real 
number values for the set of eigenstates, as the eigenvalues. But the representation loses
the variable nature of the dynamic quantity. The discussion above resolves that problem.
The DH matrix value for the reference state as an eigenstate is then exactly $\sigma^{\ssc 3}$,
while the value for the other eigenstate, as $\ket{1}$ is  
{\tiny $\left(\begin{array}{cc}
     - 1 & 0 \\   0 & 1
      \end{array}  \right)$}.                     In the other representation 
of the noncommutative values as a sequence of complex numbers we discussed in the 
introduction section, such a sequence for an eigenstate would have the first term as the 
eigenvalue and the rest being zero (see the explicit form given in Ref.\cite{097}), with the 
zeros still playing an important role in the properties of the noncommutative value as an 
element of a noncommutative algebra. One way or another, only that single eigenvalue 
cannot fully represent the value of the quantum observable. Each eigenstate as 
a normalized vector is defined physically up to a phase factor. Such phase factors play no 
role in the naive matrix representation. The ordering of the eigenstates, seen in the ordering 
of the eigenvalues along the diagonal of the matrix, of course, matters. The matrix is 
a conventional defined DH matrix value. But that is not enough. The set of basis vectors is 
not specified completely unambiguously. The relative phase factor here is related to the 
conventional defined DH matrix value of the other independent observable we take as 
$s^{\ssc 1}$ here. The choice of the set of DH matrix values for the reference state for
 $s^{\ssc 3}$ and $s^{\ssc 1}$ as $\sigma^{\ssc 3}$ and $\sigma^{\ssc 1}$ leaves no
 ambiguity and uniquely gives the DH matrix values for all observables. That allows the 
 comparison of the values not only among different observables for the same state, but 
 also among those of any states. 
 
 To summarize, we have explicitly written down a conventionally taken scheme that 
 unambiguously assigns a matrix with explicit matrix elements to each observable in any 
 particular state. The assignment preserves the mathematical relations among observables 
 in the values for the same state and also gives a consistent picture of how the values of 
 the same observables change as the state changes. That is a complete picture of how the 
 system of $2\times 2$ matrices serves a system of numbers useful for modeling the values 
 of the physical quantities as dynamic variables for the qubit system. 
       
\section{Concluding Remarks}
We have presented above a careful discussion of all the notions involved in the seeing matrices 
as the q-numbers Dirac wanted for modeling values of physical quantities as best described 
in the quantum theory, at least for any system with an $N$-dimensional Hilbert space. The last 
section in particular illustrates that all fully explicitly for a qubit, that can be adopted for 
a system of qubits. The basic part of it is just what has been taken as the matrix representation 
of the obervables. Only that our picture here should be the right physics picture. Observables 
are dynamic variables. An explicit one-to-one matrix representation of the observable algebra 
is really a conventional scheme of assignment of q-numbers as $N\times N$ matrices to their 
(DH matrix) values for a system with an $N$-dimensional Hilbert space, with the exact matrix 
representation as the set of values for a reference state. Our presentation restores the nature 
of the observables as dynamic variables, having state-dependent values. The scheme gives 
a one-to-one homomorphism of the observable algebra into the q-numbers. From the theoretical 
and mathematical point of view, such a q-number value encodes the full information the 
quantum theory presents about the corresponding observable for the specific state. Complete 
information about a quantum system can be encoded in the set of local basic observables 
independent of whether the state is an entangled one \cite{DH,097}. One can look at the
basic observables as a set of coordinates for the quantum phase space, with a state to be
specified by their values. That is an exact analog of the classical case, except that the phase
space then should be seen as having a noncommutative geometry \cite{078}. The observables
are the physical quantities. Hence the noncommutative picture is the physical picture.

Values of physical quantities are always only conventionally defined. The actual value of one 
such quantity has no true meaning. Only the relations among such quantities, conveniently
taken as the relation of each such value to a chosen standard value(s) characterize its
meaning. There are reference frame issues involved too. The full conventional scheme
in classical physics is well-known. The main task of this article is to clarify the content of
such a convention necessary for quantum physics, to be seen as Dirac suggested, and we
see most appropriate at least from the theoretical and conceptual perspectives, as q-number 
physics, or matrix-valued physics. Apart from having the noncommutative (q-)numbers, the 
matrices, to be used to model mathematically the notion of values for the physical quantities, 
quantum physics is essentially the same as classical physics. That is the original idea of 
Heisenberg, that we need mostly only to think about the mathematical properties of the
dynamic quantities differently, not as real number-valued variables.

Let us say a few more words about the $N>2$ cases. In Ref.\cite{097}, we have addressed
how one may define the DH matrix values of a generic quantum system unambiguously. The 
only thing missing in the latter paper is the explicit identification of the values as noncommutative 
number values and the discussion of what is the conventional features involved. In Ref.\cite{097}, 
it has already been emphasized that a proper $U_{\!\ssc \phi}$ for such a scheme should contain 
only the parameters or the set of independent coordinates needed to fully specify a generic state. 
A simple scheme as a generalization of what we have in the last section to the case $N>2$ can 
easily be appreciated. Explicitly, for a simple $N=3$ system, one can take 
 \bea
  U_{\!\ssc \phi} =\left(\begin{array}{ccc}
  c & - \bar{s}  & 0\\
  s & \bar{c} & 0 \\
  0 & 0 & 1
  \end{array}  \right)
  \left(\begin{array}{ccc}
    1 & 0 & 0 \\
    0 & c' & - \bar{s}'  \\
    0 & s' & \bar{c}'  
    \end{array}  \right)
  \eea
 which corresponds to the state $\ket\phi = c \ket{0} + sc' \ket{1}  + ss' \ket{2}$. And similarly 
 for a larger $N$. However, for a multipartite composite system, the observable algebra as 
 a tensor product has a number of generators smaller than that of the full matrix algebra for 
 a generic system with Hilbert space of the same dimension $N$, as that of the composite 
 system. The algebra of DH matrix values for any particular state then is only a nontrivial 
 subalgebra of the $N\times N$ matrices as a noncommutative number system. For a composite 
 system, It is more interesting to take  a $U_{\!\ssc \phi}$ that coordinates the states separating 
 the parameters that describe entanglement from others. Other than that, the conventional 
 scheme of assignment of the DH matrix values can naturally be taken from that of the individual 
 part, based on the local basic observables adapted as operators on the composite system 
 Hilbert space. Interested readers may consult a description of the DH matrix value for a 
 two-qubit entangled state given in Ref.\cite{097}, for example. Of course, that is only about how 
 we choose to describe things for a generic state. That is useful for analyzing some theoretical 
 issues about how some physical properties change as the state changes. For the DH matrix 
 values of observables for any specific state, we have an explicit numerical $U_{\!\ssc \phi}$. 
 The description of the noncommutative number values of the system adopted for each qubit in 
 a system of qubits as a full description of, at least the theoretical, quantum information in the 
 system is what motivated the study by Deutsch and Hayden \cite{DH}, though the authors did
 not use the term noncommutative number values nor address the issues in the general setting 
 of quantum physics. How to use them in practical studies and applications of quantum 
 information is the challenge at hand.
 
 Mathematically, the system of $N\times N$ matrices for each $N$ can be embedded into one 
 with a larger $N$, and we have larger and larger noncommutative number systems. From the 
 point of view of fundamental physics taking particles as the basic `elementary' systems, the 
 basic observables for a particle are the position and momentum observables as generators of 
 the observable algebra. One has to take $N$ to countable infinity. Seeing the position
 and momentum observables as the noncommutative coordinates of the quantum phase 
 space \cite{078} is a very intuitive idea. The expectation value functions, taken as the 
 starting point for our earlier notion of the noncommutative values of the observables, 
 also serve the purpose of giving a picture of the observables as functions of the usual 
 real/complex number geometric picture of the phase space as the (projective) Hilbert space. 
 That is a classical-like picture of the observable algebra for the symplectic geometry, except
 with a noncommutative product  \cite{CMP}. Matching that to the operator algebra picture, 
 one can see the Schr\"odinger picture and the Heisenberg picture as about Hamiltonian 
 dynamics on the same symplectic geometry, the former describes that with the infinite 
 number of real/complex number coordinates while the latter describes that with the small 
 number of noncommutative coordinates \cite{078}. The noncommutative number values of 
 the latter coordinates then give a specification of a point in the noncommutative symplectic 
 geometry. A generic scheme of local coordinate picture for noncommutative geometry may 
 be possible. That would be a true geometric picture to complement the mostly algebraic 
 descriptions in mathematical subject.  It may also be interesting to consider, from the point 
 of view of mathematics, the possibility of generalizing the noncommutative number systems 
 to one of a skew field. Another direction is to consider generalizing the notions of vector space 
 and algebra to one with noncommutative numbers as scalars, and a full noncommutative version 
 of Hilbert's Nullstellensatz may be possible. 
 
We want to emphasize again that having the evaluation maps as algebraic homomorphisms 
is physically important. The theoretical relations among physical quantities can only be checked 
through checking the relations among their values, {\em i.e.} through checking experimentally 
the relations the theory predicts among all information of the observables theoretically available. 
Not being able to do that means that we cannot fully confirm the theory. For example, when the 
theory says an observable is the product of two others, it is the values of the latter for any specific 
state agreeing with that for the product that verifies it. With only the notion of the eigenvalue 
distributions in projective measurements, such a verification cannot be achieved. The success 
of quantum mechanics as a theory certainly could not be established only by obtaining expectation 
values of observables, still less by looking only at eigenvalue results of individual projective 
measurements for each observable. Experimental accessibility is of course important. However, 
how directly that may be could be mostly a question about our practical ingenuity. In our previous 
studies \cite{088,079}, we have discussed how, to the extent that quantum state can be 
experimentally determined \cite{L}, our earlier notion of the noncommutative values can be
determined through enough experimental efforts. In principle, that can be done up to any required
precision. It could be just determining real number quantities, much like determining the full
statistical distribution of eigenvalue outcomes in projective measurements through determining 
the moments. The latter is more or less what experimental physicists have been doing when
checking the quantum theory, though one may be talking about it as the `probabilities' of the
individual outcomes. It is easy to see that the same holds for the DH matrix value. What is more
interesting and truly challenging is if one can find some ways the full value more directly, starting
with the question of how to implement a reference system for defining the convention adopted
in fixing the values. The DH matrix values as a sensible notion of noncommutative numbers that
can be dealt with in a generic manner independent of the physical state is particularly important 
for the purpose. The fact that the noncommutative numbers do not form a well-ordered set is 
certainly going to make the task more complicated than simply calibrating a simple output scale 
with noncommutative numbers, even if one finds a way to do that. Yet, again, we think that is 
about treating quantum information as quantum information. Physical measurement is about 
extracting information about the physical quantity we are interested in experimentally. As the 
physical information is, at the fundamental level quantum in nature, we are dealing with quantum 
information. A noncommutative number gives, mathematically, a direct description of an individual 
piece of quantum information. The Deutsch and Hayden paper is effectively the first step to use 
that in quantum information science. This article aims at a full clarification of the notion and 
theoretical issues involved, hoping to lay a firm ground for practical efforts in its applications.

\section*{Acknowledgment}
The author is partially supported by research grant 
number   112-2112-M-008-019 of the NSTC of Taiwan

\end{document}